\documentclass[a4paper,pre,aps,showkeys,showpacs,twocolumn,amsmath,amssymb,superscriptaddress]{revtex4}
\usepackage[dvipdf]{color}
\usepackage{graphicx}
\usepackage{dcolumn}
\usepackage{bm}
\usepackage{times}

\begin{document}

\title{Scattering from generalized Cantor fractals}
 
\author{A. Yu. Cherny}
\affiliation{Joint Institute for Nuclear Research, Dubna 141980, Russian Federation}

\author{E. M. Anitas}
\affiliation{Joint Institute for Nuclear Research, Dubna 141980, Russian Federation}
\affiliation{Horia Hulubei National Institute of Physics and Nuclear Engineering, 
RO-077125 Bucharest-Magurele, Romania} 
 
\author{A. I. Kuklin}
\affiliation{Joint Institute for Nuclear Research, Dubna 141980, Russian Federation}
 
\author{M. Balasoiu}
\affiliation{Joint Institute for Nuclear Research, Dubna 141980, Russian Federation}
\affiliation{Horia Hulubei National Institute of Physics and Nuclear Engineering, 
RO-077125 Bucharest-Magurele, Romania}
 
\author{V. A. Osipov}
\affiliation{Joint Institute for Nuclear Research, Dubna 141980, Russian Federation}

\date{\today}
\begin{abstract}
We consider a fractal with a variable fractal dimension, which is a generalization of 
the well known triadic Cantor set. In contrast with the usual Cantor set, the fractal dimension is 
controlled using a scaling factor, and can vary from zero to one in one dimension and from zero 
to three in three dimensions. The intensity profile of small-angle scattering from the 
generalized Cantor fractal in three dimensions is calculated. The system is generated by a set 
of iterative rules, each iteration corresponding to a certain fractal generation. Small-angle scattering is considered from monodispersive sets, which are randomly 
oriented and placed. The scattering intensities represent minima and maxima superimposed 
on a power law decay, with the exponent equal to the fractal dimension of the 
scatterer, but the minima and maxima are damped with increasing polydispersity of the 
fractal sets. It is shown that for a finite generation of the fractal, the exponent 
changes at sufficiently large wave vectors from the fractal dimension to four, the 
value given by the usual Porod law. It is shown that the number of particles of which 
the fractal is composed can be estimated from the value of the boundary between the 
fractal and Porod regions. The radius of gyration of the fractal is calculated 
analytically.
\end{abstract}


\maketitle

\section{Introduction}
\label{sec:intro}

Small-angle scattering (SAS; X-rays, light, 
neutrons)~\cite{svergun87:book,zemb02:book,glatter82:book} is one of the most important 
investigation techniques for the structural properties of fractal 
systems~\cite{pfeifer83,mandelbrot83:book,pfeifer89:book,peitgen04:book} at the nanometer 
scale. This technique yields the Fourier transform of the spatial density distribution 
of a studied system within a wave vector range from  $10^{-4}$ to 
$10^{-1}$\AA$^{-1}$ and is thus very effective for object sizes from $10$ to 
$10^4$\AA~\cite{svergun87:book,zemb02:book,glatter82:book}. Since only a finite range in 
wave vector space is available experimentally, the interpretation of the results 
requires theoretical models.  

In the case of non-deterministic (random) fractals, different models for mass and 
surface fractals \cite{bale84,sinha84:book,martin85,freltoft86,martin86,hurd87} have 
been successfully developed and applied to a variety of materials~\cite{malcai97}. However, for 
deterministic fractals, only a few attempts have been made
made~\cite{schmidt86,kjems86:book,schmidt91}. This is probably due to the technological 
limitations encountered until just a few years ago in the preparation of fractal-like 
structures. Modern experimental techniques for obtaining deterministic fractal 
systems have now been developed \cite{mayama06,tkmsh04,cnar08}. Small-angle neutron 
scattering (SANS) from self-assembled porous silica materials \cite{mayama06} was 
carried out by Yamaguchi \textit{et al.} (2006)~\cite{yamaguchi08}. A molecular Sierpinski hexagonal gasket, the deterministic fractal polymer, was also obtained by chemical methods \cite{newkome06}. 

The main indicator of fractal structure is the power law behaviour of the scattering 
curve
\begin{equation}
I(q)\sim q^{-\alpha},
\label{scattexp}
\end{equation} 
where $\alpha$ is the power law scattering exponent. $I$ is the diffracted intensity and $q=\frac{4\pi}{\lambda}\sin\theta$, where $2\theta$ is the scattering angle and $\lambda$ is the incident wavelength. The expnent $\alpha$ carries information regarding 
the fractal dimension of the scatterer 
\cite{bale84,sinha84:book,martin85,freltoft86,martin86,hurd87,sorensen99}: $\alpha = D$ 
for mass fractals and $\alpha = 6-D$ for surface fractals. One can write down an even 
more general expression \cite{pfeifer02} for the exponent in a two-phase geometric 
configuration, where one phase is a mass fractal of fractal dimension $D_\mathrm{m}$ 
containing pores of fractal dimension $D_\mathrm{p}$. In addition, the boundary surface 
between the phases also forms a fractal of dimension $D_\mathrm{s}$. The exponent then 
reads $\alpha = 2(D_\mathrm{m} +D_\mathrm{p})-D_\mathrm{s}-6$. In the particular case of a 
mass fractal, we have $D_\mathrm{s}=D_\mathrm{m}<3$ and $D_\mathrm{p}=3$, while for a 
surface fractal $D_\mathrm{m}=D_\mathrm{p}=3$ and $2<D_\mathrm{s}<3$.

The construction of non-random fractal models assumes the presence of an initial set 
(initiator) and a generator (iterative operation). Usually, an initiator is
divided into subparts. Some of them are then removed according to an iterative 
rule and the process is repeated for each remaining part.

In contrast with the simple power law behaviour above [equation~(\ref{scattexp})], the scattering from 
non-random fractals, such as the Menger sponge, fractal jack~\cite{schmidt86} or other 
surface-like type~\cite{schmidt91}, shows a successive superposition of maxima and minima 
decaying as a power law, which can be called a generalized power law. This is due to 
spatial order in non-random fractals: the scattering curve $I(q)$, being the Fourier 
transform of the pair distribution function, oscillates  by reason of singularities in 
this function or its derivatives. If there is disorder in the system then the 
singularities are smeared out, as in the case of random fractals, the behaviour of which follows 
the power law of equation (\ref{scattexp}). On the other hand, real physical systems always exhibit a specific kind of disorder, polydispersity.  Below, we show that when the polydispersity 
of non-random fractals is taken into account, these local oscillations cancel out with 
increasing width of the distribution function. This is consistent with the results 
obtained for a similar fractal, the Menger sponge, by averaging over a Schulz distribution function 
\cite{schmidt95:book}.
 
 The existing models \cite{schmidt86,schmidt91} assume a given fractal dimension, which 
imposes some restrictions on applying the models to real systems. In this paper, a 
generalization of Cantor sets is considered that allows us to control the fractal 
dimension using a scaling factor. The scattering intensity of randomly oriented 
generalized Cantor sets is obtained analytically for mono- and polydisperse 
systems. Our method is based on the approach that was successfully employed in some 
aspects for other similar systems of non-random fractals such as the Menger sponge 
\cite{schmidt86}. 

Numerous examples of the fractal-like behaviour were obtained experimentally in 
collaboration with one of the authors of this paper (AIK) for different kind of 
objects, namely soils \cite{fedotov06,fedotov07,fedotov07a}, biological objects 
\cite{lebedev08} and nanocomposites \cite{dokukin07}. Only three parameters  are 
extracted from the fractal scattering intensity: its exponent $\alpha$ [see 
equation~(\ref{scattexp})], and the edges of the fractal region in $q$-space, which appear as 
"knees" in the scattering line on a logarithmic scale. Other parameters are not 
usually obtained from small-angle neutron scattering (SANS) curves.

If some features of the fractal structure are available from other considerations, one 
can construct a model and obtain additional information on the fractal parameters. 
Note that a real physical system does not possess an infinite scaling effect and the 
number of iterations is always finite. In this paper we calculate the scattering from 
generalized Cantor sets for a finite number of iterations. As discussed below, many 
features of the scattering are quite general; in particular, one can estimate the number 
of particles from which the fractal is formed.

\section{Construction of the generalized Cantor set} 
\label{sec:set}

The well known Cantor set is created by repeatedly deleting the open central third of a 
set of line segments. One can construct various generalizations of the set by, say, cutting the segments into a different number of intervals or choosing another scaling factor 
(see, e.g. \cite{gouyet96:book}). Below, we explicitly describe a three-dimensional 
generalization of the Cantor set, which is used for studying the scattering in the subsequent 
sections. 

\begin{figure}
\centering
\includegraphics[width=\columnwidth,clip=true]{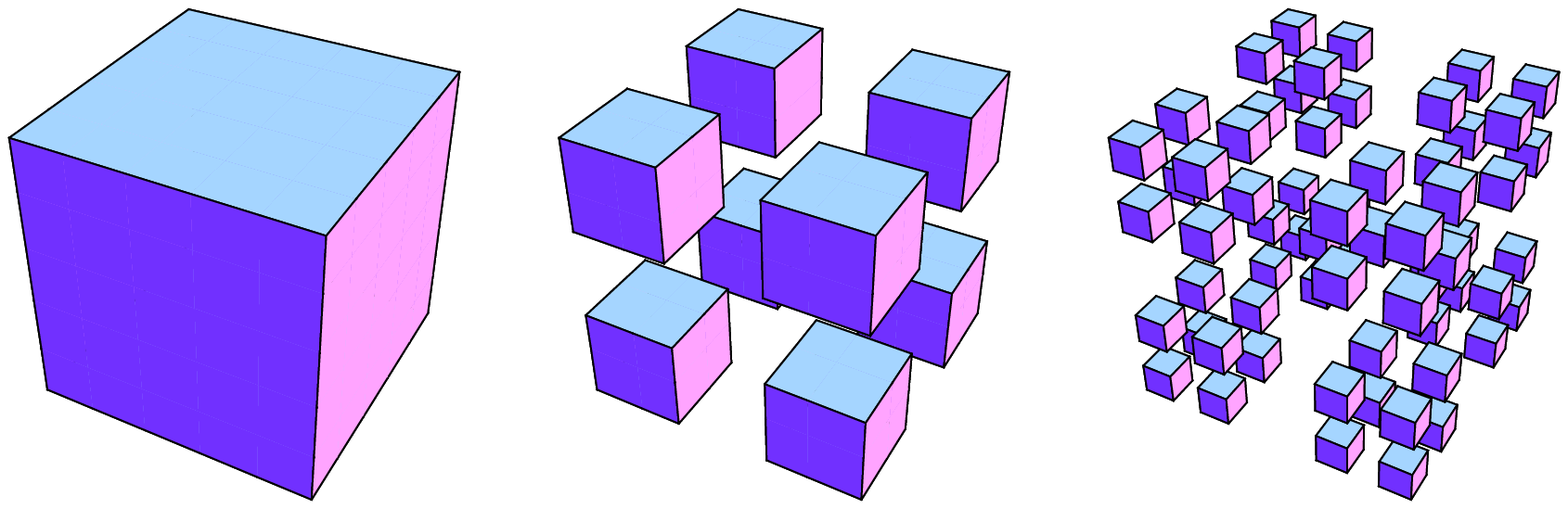}\\
\includegraphics[width=\columnwidth,clip=true]{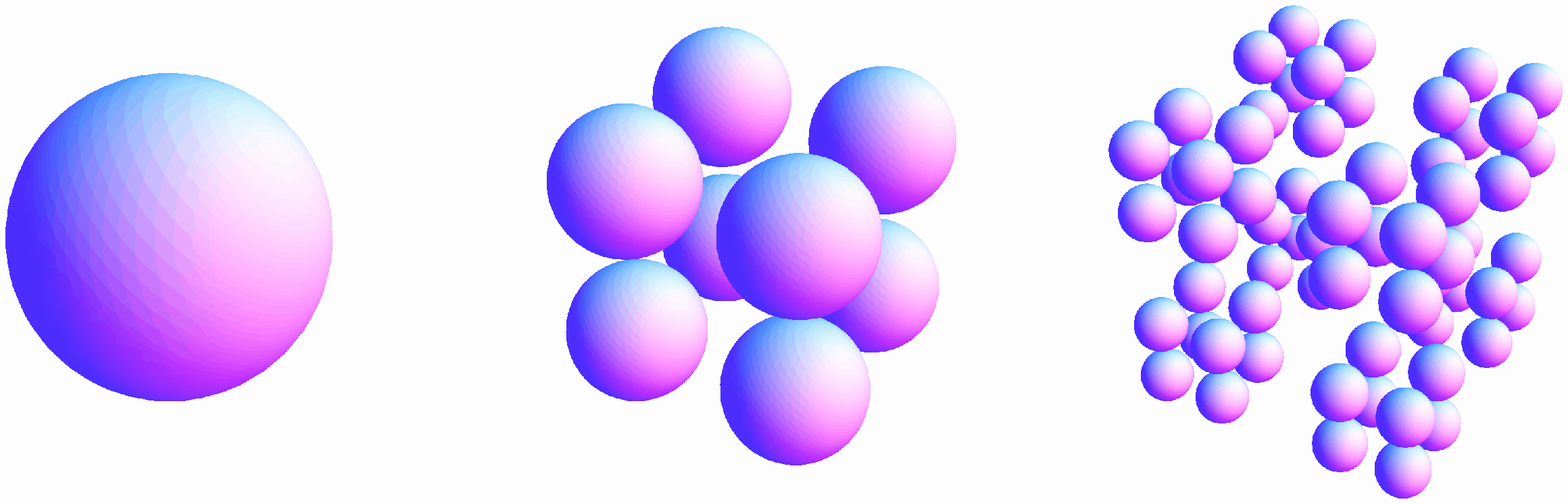}
\caption{
Zero, first and second iterations (approximants) for the 
generalized Cantor set, composed of cubes (upper panel) and balls (lower panel).}
\label{fig:3dcantorset}
\end{figure}

The generalized Cantor set in three dimensions can be constructed from a homogeneous 
cube by iterations, which we will call approximants. The zero-order iteration 
(initiator) is the cube itself with side length $l_0$, which can be specified in 
Cartesian coordinates as a set of points obeying the conditions $-l_0/2\leqslant 
x\leqslant l_0/2$, $-l_0/2\leqslant y\leqslant l_0/2$, $-l_0/2\leqslant z\leqslant 
l_0/2$. The first iteration removes all points from the cube with coordinates 
$-\gamma l_0/2< x< \gamma l_0/2$ or $-\gamma l_0/2< y< \gamma l_0/2$ or $-\gamma l_0/2< 
z< \gamma l_0/2$. The dimensionless scaling factor $\gamma$ can take any value 
between zero and three. Thus, the initial cube is divided into 27 parts, the eight 
cubes are left in the corners, with edge length $(1-\gamma)l_0/2$, and the 
19 parallelepipeds are removed (see Fig.~\ref{fig:3dcantorset}, upper panel).

In order to obtain the second approximant, one should  do the same with each of the 
eight cubes, thus leaving 64 cubes of side length $(1-\gamma)^2l_0/4$. 
It is not difficult to see that the $m$th approximant to the three-dimensional Cantor 
set is composed of $N_m=8^m$ cubes of the side length
\begin{equation}
l_m=(1-\gamma)^ml_0/2^m. \label{ln}
\end{equation}
The generalized Cantor set is obtained in the limit $m\to\infty$.

The Hausdorff dimension~\cite{mandelbrot83:book} of the set can be determined from the 
intuitively apparent relation $N_m\sim (l_0/l_m)^D$ for a large number of iterations, which 
yields
\begin{equation}
D=\lim_{m\to\infty}\frac{\ln N_m}{\ln (l_0/l_m)}=-\frac{3\ln 2}{\ln \beta_\mathrm{s}},
\label{dimension}
\end{equation}
where the scaling factor $0<\beta_\mathrm{s}<1$ for each iteration is defined as
\begin{equation}
\beta_\mathrm{s}\equiv(1-\gamma)/2.
\label{betas}
\end{equation}
In the same manner, it can be shown that the fractal dimension of the surface coincides with 
that of the bulk structure. We then have a mass fractal, the dimension of which can be varied 
between zero and three by changing the parameter $\gamma$. 

\begin{figure}
\centering
\includegraphics[width=.4\columnwidth,clip=true]{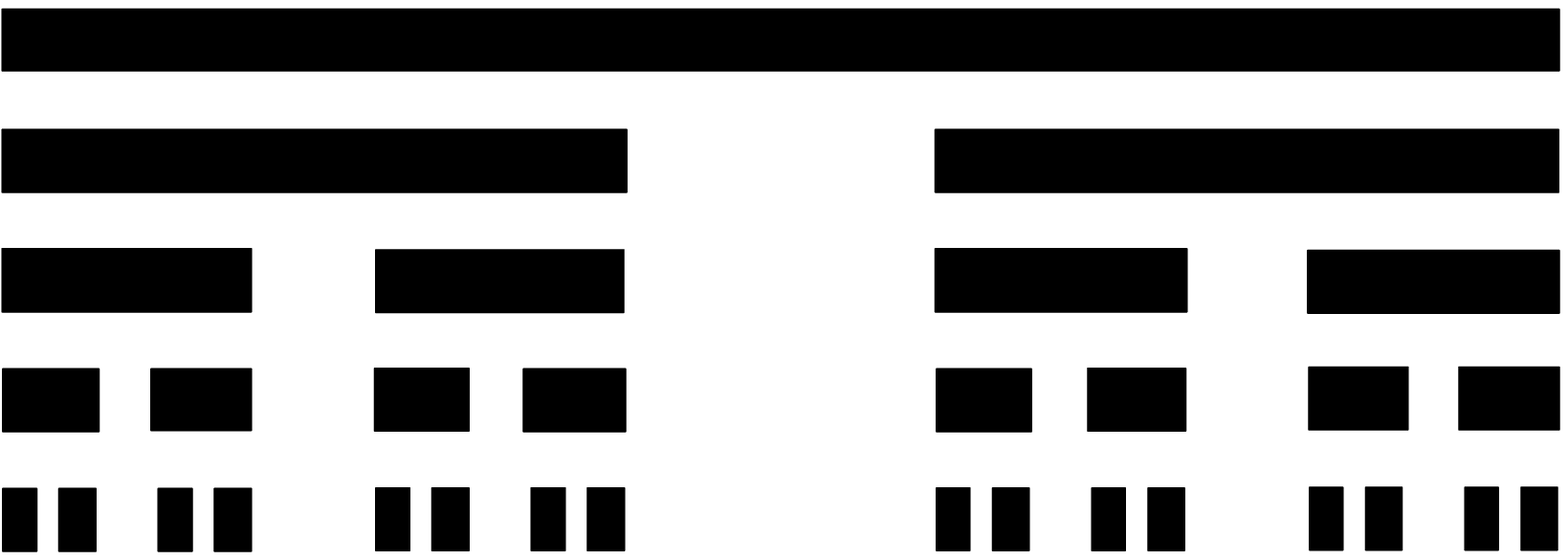}\hspace{.1\columnwidth}
\includegraphics[width=.4\columnwidth,clip=true]{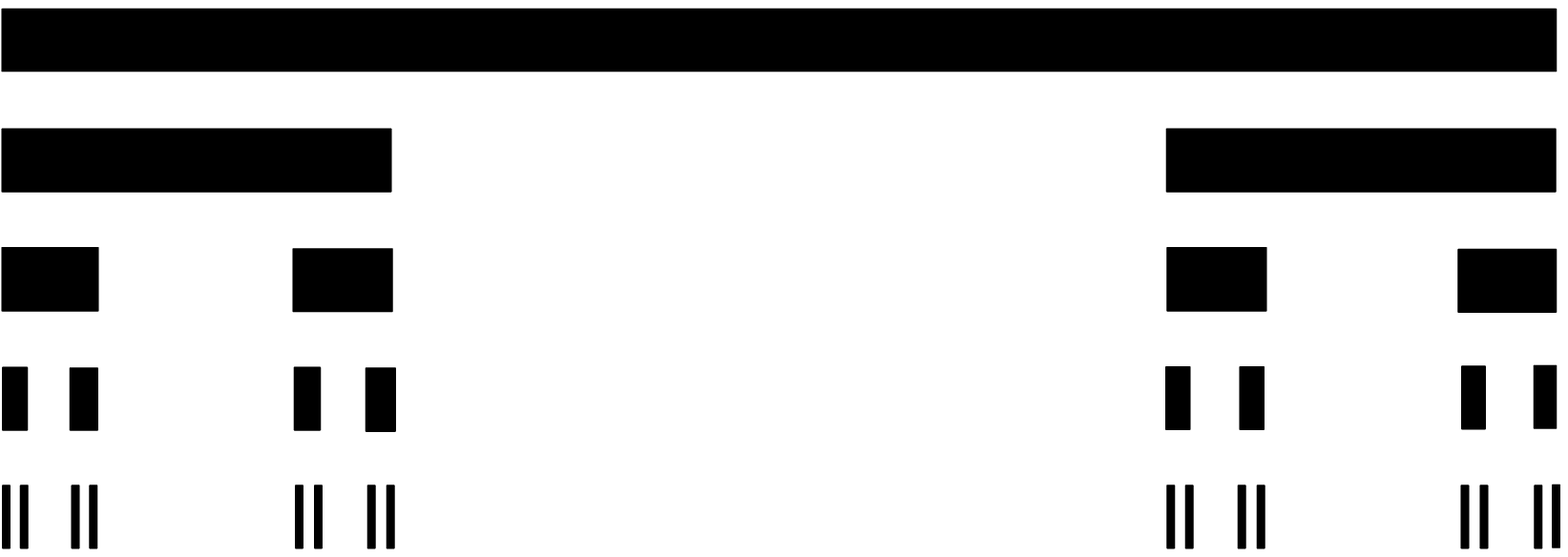}
\caption{The first five iterations for the one-dimensional 
generalized Cantor set with scaling factors $\gamma=0.2$ (left) and $\gamma=0.5$ 
(right). It is generated by removing at the $m$th iteration the central interval of length 
$\gamma^m l_0$ from each remaining segment. The fractal (Hausdorff) dimension of the 
one-dimensional generalized Cantor set is equal to $-\ln 2/\ln \beta_\mathrm{s}$, where 
$\beta_\mathrm{s}$ is given by equation~(\ref{betas}). At $\gamma=1/3$ we obtain the usual 
Cantor set with fractal dimension $\ln 2/\ln 3=0.6309\ldots$. }
\label{fig:1dcantorset}
\end{figure}

Note that the generalized Cantor set composed of cubes is simply the direct product 
of three one-dimensional generalized Cantor sets (Fig.~\ref{fig:1dcantorset}). This 
means that the point $\bm{r}=(x,y,z)$ belongs to the three-dimensional set if and only 
if each of the coordinates $x$, $y$ and $z$ belongs to the one-dimensional set.  The 
fractal (Hausdorff) dimension of the one-dimensional generalized Cantor set is three 
times smaller than that of the three-dimensional set [equation~(\ref{dimension})].

One can generalize the fractal construction, replacing the cubes with any solid 
three-dimensional shape and using the same scaling [equation~(\ref{betas})] for each iteration. 
For example, a ball of arbitrary radius $R\leqslant l_0/2$ in the center of the cell can 
be used as an initiator instead of the initial cube (Fig.~\ref{fig:3dcantorset}, lower panel). One can see that the fractal dimension is still given by [equation~(\ref{dimension})].

\section{Form factor of the monodisperse set}
\label{sec:mono}

\subsection{General remarks}
\label{sec:genrem}

We consider a two-phase system of particles, consisting of small pieces of ``mass". The 
masses are embedded in a solid matrix, which can be associated with the ``pores" 
discussed in Sec.~\ref{sec:intro}. The particles are also immersed in the solid matrix 
or dissolved in a solution of the same scattering length density as pores. The scattering 
length densities are $\rho_\mathrm{m}$ for the mass and $\rho_\mathrm{p}$ for the 
pores. They are determined by the relation $\rho\equiv\sum_i b_i/\Delta V$, where the
summation is over the scattering centers located in the volume $\Delta V$. The 
difference $\rho_\mathrm{m}-\rho_\mathrm{p}$ is called the scattering contrast. In what 
follows, we mean by particle the $m$th approximant to the generalized Cantor set.

The neutron scattering cross section per unit volume of a sample is given 
in~\cite{svergun87:book} as
\begin{equation}
\frac{1}{V'}\frac{d \sigma}{d \Omega}= n (\rho_\mathrm{m}-\rho_\mathrm{p})^{2} 
V^{2}\langle|F(\bm{q})|^{2}\rangle, \label{cross}
\end{equation}
where $\bm{q}$ is momentum transfer, $n$ is the number of particles per unit volume and $V$ 
is the total volume of the mass in the particle. [Note that, in the literature, the quantity $V F(\bm{q})$ is sometimes called the scattering amplitude, while
the squared and averaged quantity $V^2\langle |F(\bm{q})|^2\rangle$ is what is meant by form factor.] The normalized scattering form factor is defined as
\begin{equation}
F(\bm{q})\equiv\frac{1}{V}\int_{V}e^{-i\bm{q}\cdot\bm{r}}d\bm{r}, 
\label{form}
\end{equation}
and the symbol $\langle \ldots \rangle$ denotes the mean value over all orientations 
of the particle. Here we assume that the particles are randomly oriented in space and 
their positions are not correlated. The latter means that we neglect particle 
interactions and put the interparticle structure factor equal to 1, which is quite 
reasonable if the average distance between the particles is sufficiently large.

If the probability of any orientation is the same, then the mean value can be calculated 
by averaging over all directions $\bm{n}$ of the momentum transfer $\bm{q}=q \bm{n}$, 
that is, by integrating over the solid angle in the spherical coordinates ${q}_{x}=q 
\cos\varphi \sin\vartheta$, ${q}_{y}=q \sin\varphi \sin\vartheta$ and ${q}_{z}=q 
\cos\vartheta$ 
\begin{equation}
\langle f(q_x,q_y,q_z) \rangle\equiv\frac{1}{4\pi}\int_{0}^{\pi}d\vartheta\sin\vartheta\int_{0}^{2\pi}d
\varphi\,f(q,\vartheta,\varphi).
\label{aver}
\end{equation}

Once we know the absolute values of the intensity [equation~(\ref{cross})], the concentration of 
fractals and the contrast, we can obtain the fractal volume from the scattering at 
zero momentum. 

\subsection{An analytical formula for the fractal form factor}

One can easily obtain an explicit analytical formula for the form factor [equation~(\ref{form})] of the 
$m$th approximant. A similar method for obtaining the form factor of non-random fractals was 
used in~\cite{schmidt86} (see also a generalization in~\cite{lidar96}). The zeroth 
approximant is an initiator with a form factor $F_{0}(\bm{q})$. If the 
initiator is a cube of edge $l_0$ we have $F_0=F_\mathrm{c}$, where
\begin{equation}
F_\mathrm{c}(\bm{q})=
\frac{\sin(q_{x}l_{0}/2)}{q_{x}l_{0}/2}\frac{\sin(q_{y}l_{0}/2)}{q_{y}l_{0}/2}\frac{\sin(q_{z}l_{0}/2)} 
{q_{z}l_{0}/2}.
\label{fcube}
\end{equation}
In the case of the ball of radius $R$, we have $F_0=F_\mathrm{b}$ with
\begin{equation}
F_\mathrm{b}(\bm{q})=3\frac{\sin(q R)-q R\cos(q R)}{(q R)^3},\quad q=|\bm{q}|.
\label{fball}
\end{equation}
To obtain the first-order formula, one can use three simple properties of the the 
scattering form factor [equation~(\ref{form})] for a particle of arbitrary shape:\\ \noindent 
\emph{i)} If we scale the entire length of a particle as $l\to\beta l$, then 
$F(\bm{q})\to F(\beta\bm{q})$.\\ \noindent \emph{ii)} If a 
particle is translated $\bm{r}\to\bm{r}+\bm{a}$, 
then $F(\bm{q})\to 
F(\bm{q})\exp(-i\bm{q}\cdot\bm{a})$.\\
\noindent \emph{iii)} If a particle consists of two non-overlapping subsets $\mathrm{I}$ and $\mathrm{II}$, 
then $F(\bm{q}) =\big(V_I F_I(\bm{q}) +V_\mathrm{II} 
F_\mathrm{II}(\bm{q})\big)/(V_\mathrm{I}+V_\mathrm{II})$.

The first approximant consists of eight cubes (balls), which differ from the initial 
cube by the scaling factor of equation~(\ref{betas}) and the center positions of which are shifted from the 
center of the initial cube by the vectors $\bm{a}_j=\{\pm\beta_\mathrm{t}l_0, 
\pm\beta_\mathrm{t}l_0, \pm\beta_\mathrm{t}l_0\}$ with various combinations of the 
signs. Here we put by definition
\begin{equation}
\beta_\mathrm{t}\equiv(1+\gamma)/4.
\label{betat}
\end{equation}
Using the definitionof equation~(\ref{form}) and the properties \emph{i)}, \emph{ii)} and \emph{iii)}, we obtain
\begin{equation}
V_1 F_{1}(\bm{q})=\sum_{j=1}^{8}\beta_\mathrm{s}^3V_0 F_{0}(\beta_\mathrm{s}\bm{q})\exp(-i\bm{q}\cdot\bm{a}_j).
\label{f1}
\end{equation}
Here, the total volume of the first approximant is given by $V_1=8V_0\beta_\mathrm{s}^3$, and 
$V_0$ is the volume of the initial cube (ball). Writing down the sum explicitly yields
\begin{equation}
F_{1}(\bm{q})=G_1(\bm{q})F_{0}(\beta_\mathrm{s}\bm{q}),
\label{f1fin}
\end{equation}
where we introduce the function $G$ defined as
\begin{eqnarray}
G_{1}(\bm{q})\equiv\cos(q_{x}l_{0}\beta_\mathrm{t})\cos(q_{y}l_{0}\beta_\mathrm{t})\cos(q_{z}l_{0}\beta_\mathrm{t}).
\end{eqnarray}
For the second approximant we repeat the same operation on $F_{1}(\bm{q})$ and obtain
\begin{equation}
F_{2}(\bm{q})=G_{1}(\bm{q})F_{1}(\beta_\mathrm{s}\bm{q})
=G_{1}(\bm{q})G_{1}(\beta_\mathrm{s}\bm{q})F_{0}(\beta_\mathrm{s}^2\bm{q}).
\end{equation}
In the same manner, we infer the general relation
\begin{equation}
F_{m}(\bm{q})=P_m(\bm{q})F_{0}(\beta_\mathrm{s}^m\bm{q}),
\label{fm}
\end{equation} 
where
\begin{equation}
P_m(\bm{q})\equiv G_{1}(\bm{q})G_{2}(\bm{q})\ldots G_{m}(\bm{q}),
\end{equation}
and
\begin{equation}
G_{m}(\bm{q})\equiv \cos(q_{x}l_{0}\beta_\mathrm{t}\beta^{m-1}_\mathrm{s})
\cos(q_{y}l_{0}\beta_\mathrm{t}\beta^{m-1}_\mathrm{s})\cos(q_{z}l_{0}\beta_\mathrm{t}\beta^{m-1}_\mathrm{s})
\label{gm}
\end{equation}
for $m=1,2,\ldots$. We can also put, by definition, $G_0(\bm{q})=1$ and $P_{0}(\bm{q})=1$ in 
order to describe the initiator.
\begin{figure}
    \centering
\includegraphics[width=.9\columnwidth]{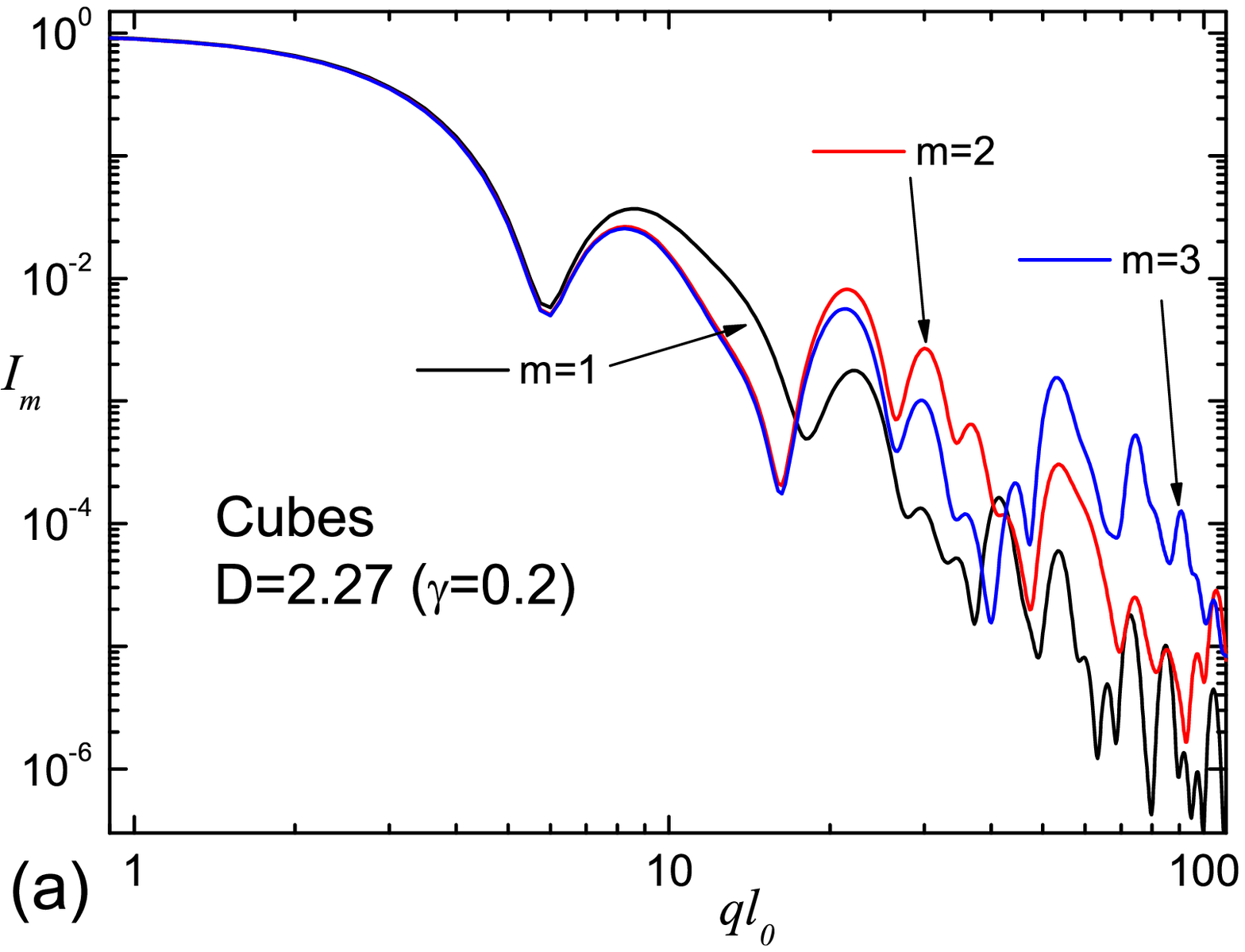}\\
\includegraphics[width=.9\columnwidth]{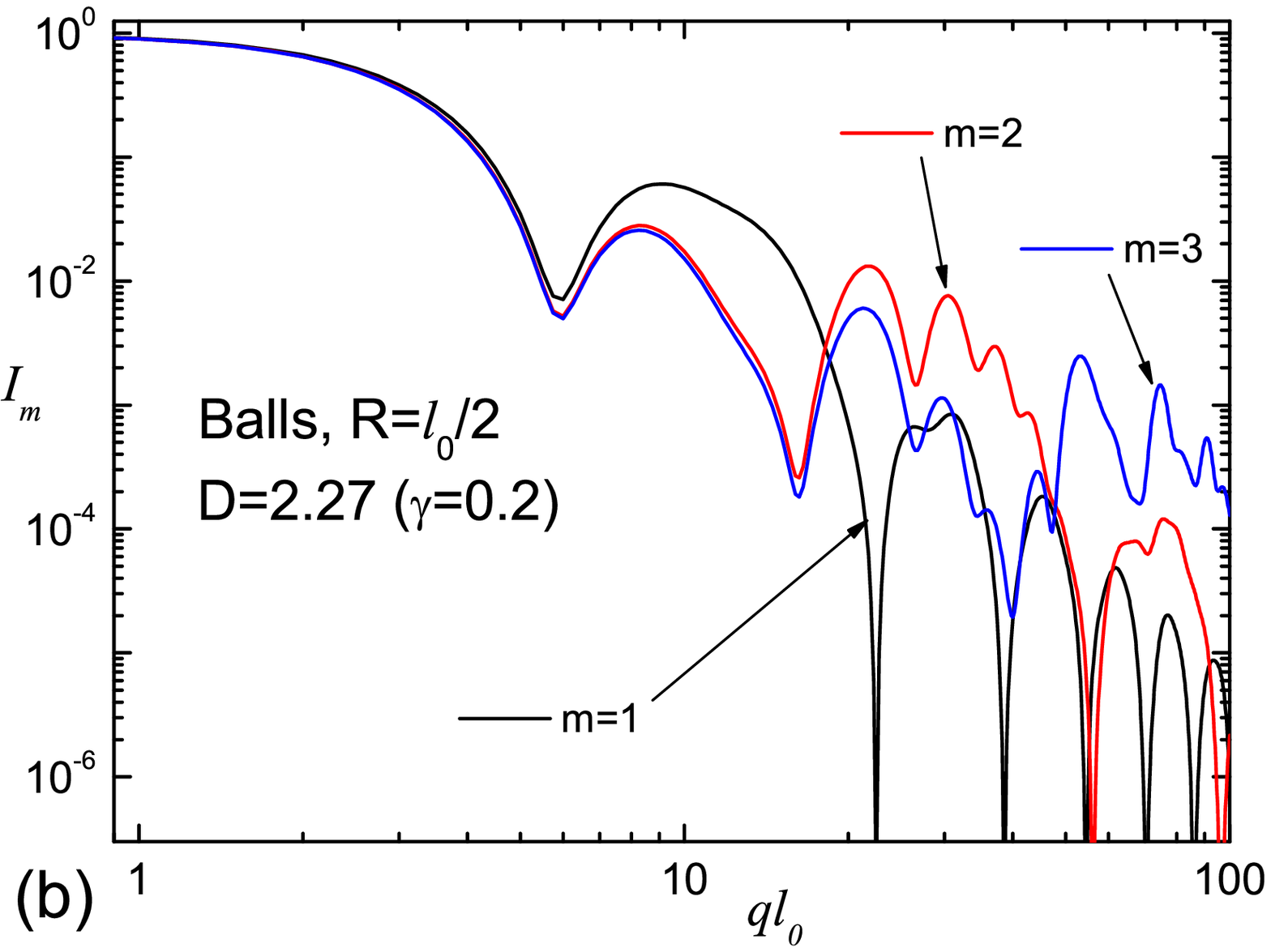}
\caption{Scattering intensities [equation~(\ref{final})] for the first three iterations of the 
generalized Cantor set composed of (a) cubes and (b) balls. The radius of the initial ball 
(initiator) is $R=l_0/2$. The intensities can be estimated as a product [equation~(\ref{intapp})] of the fractal structure factor and scaled zero iteration, shown in Fig.~\ref{fig:iter}.}
\label{fig:itermono}
\end{figure}

To obtain the cross section given by equation~(\ref{cross}), we should calculate the mean 
value of $|F_{m}(\bm{q})|^2$ over all directions $\bm{n}$ of the momentum 
transfer using equation~(\ref{aver})
\begin{align}
I_m(q l_0)\equiv\langle |F_{m}(\bm{q})|^2 \rangle,
\label{final}
\end{align}
where $m=0,1,\ldots$. Here, the form factor $F_{m}(\bm{q})$ is given by 
equations~(\ref{fm})-(\ref{gm}) in conjunction with equations~(\ref{betas}) and (\ref{betat}); the 
initial form factor $F_0$ is represented by equation~(\ref{fcube}) or (\ref{fball}). Note that 
the intensity equation~(\ref{final}) of the $m$th monodisperse appoximant depends on the 
wave vector and the initial length through the product $ql_0$ only. The results for the first 
three iterations are shown in Fig.~\ref{fig:itermono}.

\subsection{Discussion of results}

These results are in accordance with scattering from similar systems like the Menger sponge, 
displaying the same behaviour of the scattering curve~\cite{schmidt86}. Those authors first omitted the last multiplier in equation~(\ref{fm}), 
assuming that $q l_0\beta_\mathrm{s}^m \ll 1$, and then used the expansion of 
cosines that allows for reducing the intensity formula to a sum of quite simple terms. 
However, the number of terms grows exponentially on increasing the number of 
iterations, and the method becomes very time-consuming even for $m=4$. By contrast, 
using the straightforward integration in equation~(\ref{final}), we can employ the exact 
relation [equation~(\ref{fm})] for arbitrary $m$, which can yield a substantial improvment in 
accuracy. At sufficiently large $ql_0$, the main contribution to the integral comes from 
a small number of narrow and high spikes, which is typical for an interference pattern. 
Nevertheless, the integral can be evaluated even at sufficiently large values of $ql_0$, 
up to $10^4$ in a reasonable amount of time.

\subsubsection{The fractal structure factor}
\label{sec:struct}

The scattering intensity can be approximately represented as
\begin{equation}
I_m(q l_0)\simeq S_m(q)\,\langle |F_{0}(\beta_\mathrm{s}^m\bm{q})|^2 \rangle/N_m,
\label{intapp}
\end{equation}
where
\begin{equation}
S_m(q)\equiv N_m \langle |P_m(\bm{q})|^2 \rangle. \label{struct}
\end{equation}  
Here, $N_m=8^m$ is the total number of cubes (balls) in the $m$th iteration, and the 
quantity $S_m(q)$ is the fractal structure factor (see, e.g. \cite{march67:book}).
The latter is $N_m$ times larger than the intensity given by equation~(\ref{final}) with $F_0=1$, for instance when the radius of the ball tends to zero. The form factor of the ball [equation~(\ref{fball})] is isotropic, and hence relation 
(\ref{intapp}) is satisfied exactly. The structure factor carries information about the 
relative positions of the cubes (balls) in the fractal. Indeed, using the definition (\ref{form}) 
and the analytical formula for the form factor [equation~(\ref{fm})], one can obtain 
\begin{equation}
P_m(\bm{q})=\frac{1}{N_m}\sum_{1\leqslant j\leqslant N_m}
\exp[-i\bm{r}_j^{(m)}\cdot\bm{q}],
\label{pm}
\end{equation}
where $\bm{r}_j^{(m)}$ are the center-of-mass coordinates of cubes (balls). Then the 
fractal structure factor [equation~(\ref{struct})] reads
\begin{equation}
S_m(q)=1+ \frac{2}{N_m}\sum_{1\leqslant k<j\leqslant N_m}\!\! \dfrac{\sin 
q r^{(m)}_{jk}}{q r^{(m)}_{jk}},
\label{sm}
\end{equation}
where
\begin{equation}
r^{(m)}_{jk}\equiv\big|\bm{r}_j^{(m)}-\bm{r}_k^{(m)}\big|.
\label{rjk}
\end{equation}
Deriving this last relation, we use the formula
$\langle\exp(i\bm{a}\cdot\bm{q})\rangle=\sin(aq)/(aq)$, which follows from 
equation~(\ref{aver}).

\begin{figure}
    \centering
\includegraphics[width=.9\columnwidth]{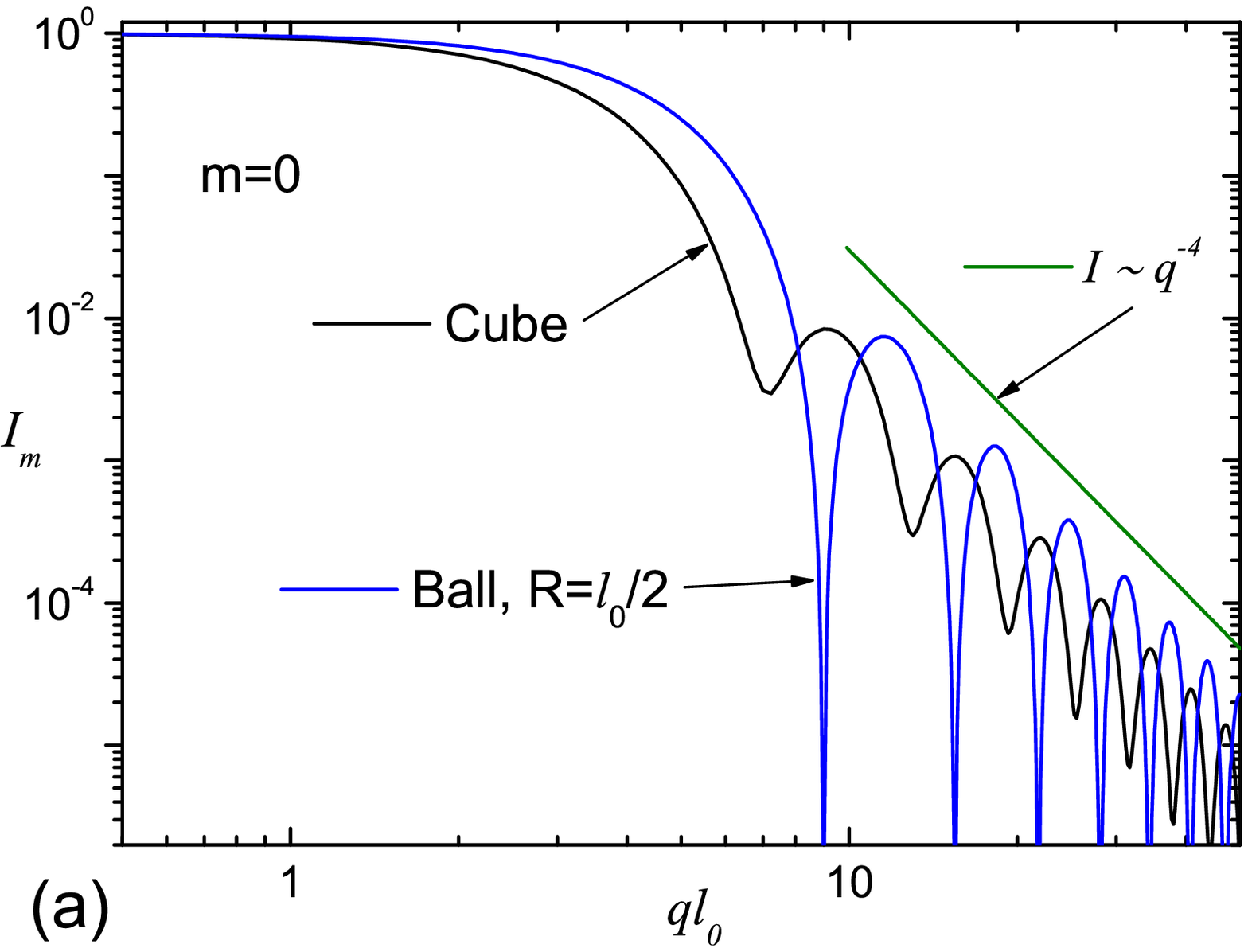}\\
\includegraphics[width=\columnwidth]{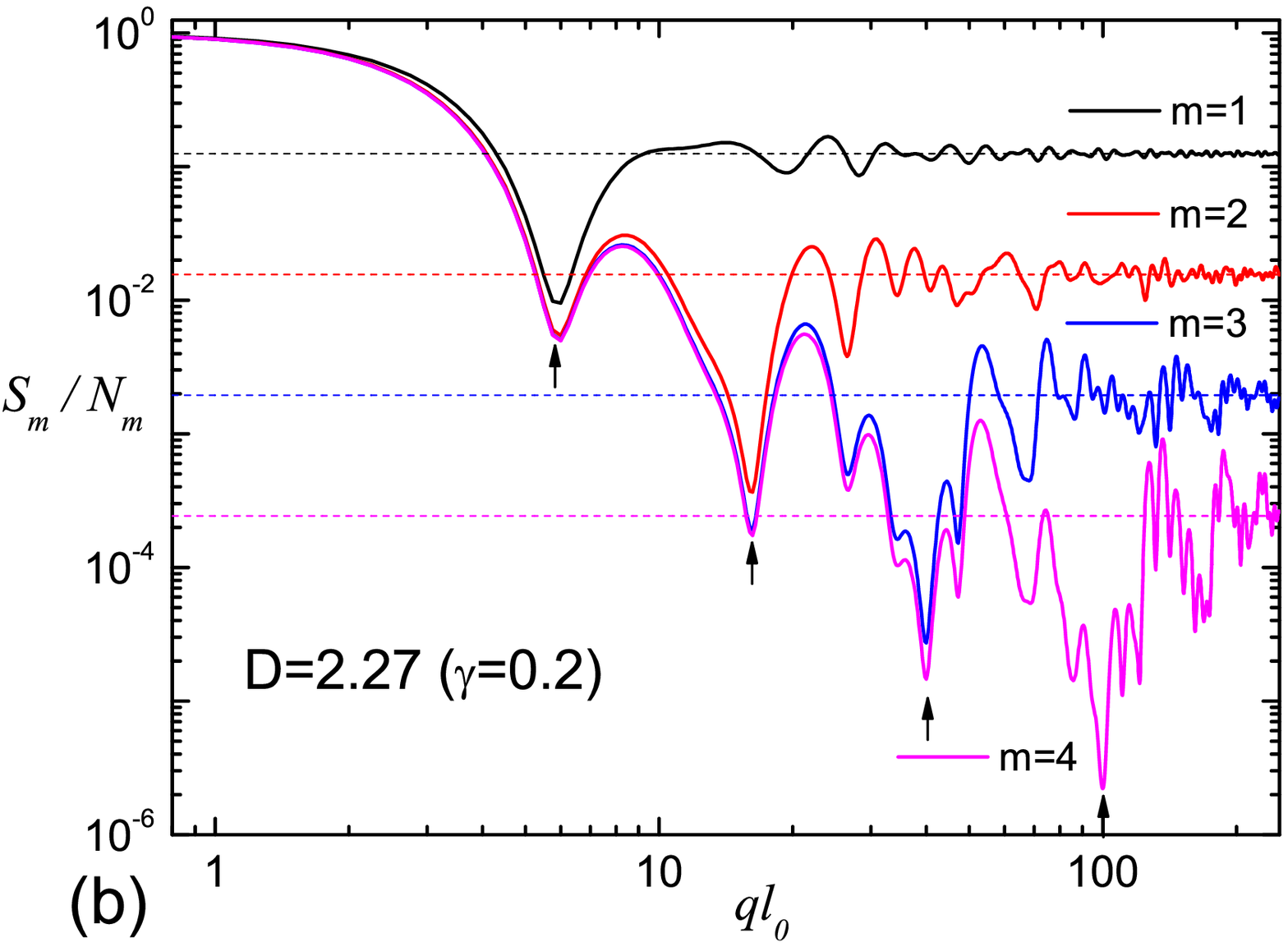}
\caption{Scattering intensities [equation~(\ref{final})] for various system geometries. (a) Zero 
iteration: intensities for a cube of edge $l_0$ and a ball of radius 
$R=l_0/2$. The positions of the maxima obey Porod`s law. (b) The structure factor of the 
generalized Cantor fractal [equation~(\ref{struct})] in units of $N_m$, where $N_m=8^m$ is the 
total number of cubes (balls) in the $m$th iteration. The diagram represents the first four 
iterations at $\gamma=0.2$, which corresponds to the fractal dimension $D\simeq 2.27$.  
The asymptote for the structure factor equals 1 when $ql_{0}\gtrsim 
1/\beta^{m}_\mathrm{s}$ (see text). The deep minima marked by 
vertical arrows can be estimated using equation~(\ref{dmin}).
    }
    \label{fig:iter}
\end{figure}

At zero momentum, relation (\ref{sm}) reads 
$S_m(0)=N_m$. At sufficiently large momentum, equation~(\ref{sm}) yields the 
asymptotic relation
\begin{equation}
S_m(q)\simeq 1.
\label{sqasymp}
\end{equation}
The underlying physics is quite clear: when the reciprocal wave vector is much smaller than 
the characteristic variance of the distance between the points then the scattering 
pattern does not ``feel" their spatial correlations. The characteristic variance is of 
the order $ql_{0}\beta_\mathrm{t} \beta^m_\mathrm{s}\sim ql_{0}\beta^m_\mathrm{s}$, and the 
asymptote is attained when 
\begin{equation}
ql_{0}\gg 1/\beta_\mathrm{s}^{m}.
\label{beyondfr}
\end{equation}

The structure factor reaches its deepest minima (see Fig.~\ref{fig:iter}b) when most of 
the objects inside the fractal interfere out of phase. This happens when the most common 
distance between the center of mass of the objects equals $\pi/q$. Taking the value of
the distance $2l_{0}\beta_\mathrm{t}\beta^{k-1}_\mathrm{s}$ (see 
Sec.~\ref{sec:set}), we obtain an estimation of the minima positions for the $m$th iteration
\begin{equation}
q_kl_0\simeq\frac{\pi}{2\beta_\mathrm{t}\beta^{k-1}_\mathrm{s}}, \quad k=1,\ldots,m.
\label{dmin}
\end{equation}
One can see that the structure factor has as many minima as the number of iterations.

\subsubsection{Generalized power law}

As was noted in Sec.~\ref{sec:intro}, rigid spatial correlations between points in 
a non-random fractal yield singularities in the form of $\delta$-functions, which leads 
to oscillations in the structure factor (Fig.~\ref{fig:iter}b). The weight of each 
$\delta$-functions follows the power law with respect to the distance between points 
$r^{(m)}_{jk}$, as determined by the fractal dimension. As a consequence, the scattering 
curve shows groups of maxima and minima which are superimposed on a monotonically 
decreasing curve proportional to some power of the absolute value of the scattering vector 
(generalized power law). Short-range correlations govern the long-range wave vector 
oscillations and \textit{vice versa}. It is essential that the oscillations are not damped on 
increasing the number of iterations, because this influences only the short-range 
correlations and hence the scattering behaviour at large momentum. 

An advantage of the model considered here is the explicit analytical expressions 
(\ref{fm})-(\ref{gm}), which allow us to estimate easily the fractal region where the 
generalized power law aplies. If the cosine argument in $G_{m+1}$ is much smaller 
than 1, then a further increasing in $m$ does not lead to an essential correction and 
the $m$th iteration reproduces the intensity that the ideal Cantor fractal would give at 
this argument. Thus, the $m$th iteration works well within the region 
$ql_{0}\beta^m_\mathrm{s}\ll 1$, where we use the estimation $\beta_\mathrm{t}\sim 1$. 
Such behaviour can be seen in  Fig.~\ref{fig:itermono}.

On the other hand, within the Gunier region $ql_0\lesssim 1$ the intensity is very close 
to 1 and we have a plateau in the logarithmic scale. The generalized power-law 
behaviour is then observed in the region
\begin{equation}
1\ll ql_0\ll 1/\beta^m_\mathrm{s}.
\label{powerlaw}
\end{equation}
This estimation indicates two typical length scales important for a fractal, its size 
$l_0$ and the characteristic distance inside the $m$th iteration $l_0\beta^m_\mathrm{s}$ 
\cite{schmidt91}. In the fractal region $F_0(\beta^m_\mathrm{s}\bm{q})\simeq 1$, and we obtain from
equation~(\ref{intapp})
\begin{align}
I_m(q l_0)\simeq S_m(q)/8^m.
\label{ifracreg}
\end{align}

\begin{figure}
\centering
\includegraphics[width=.95\columnwidth]{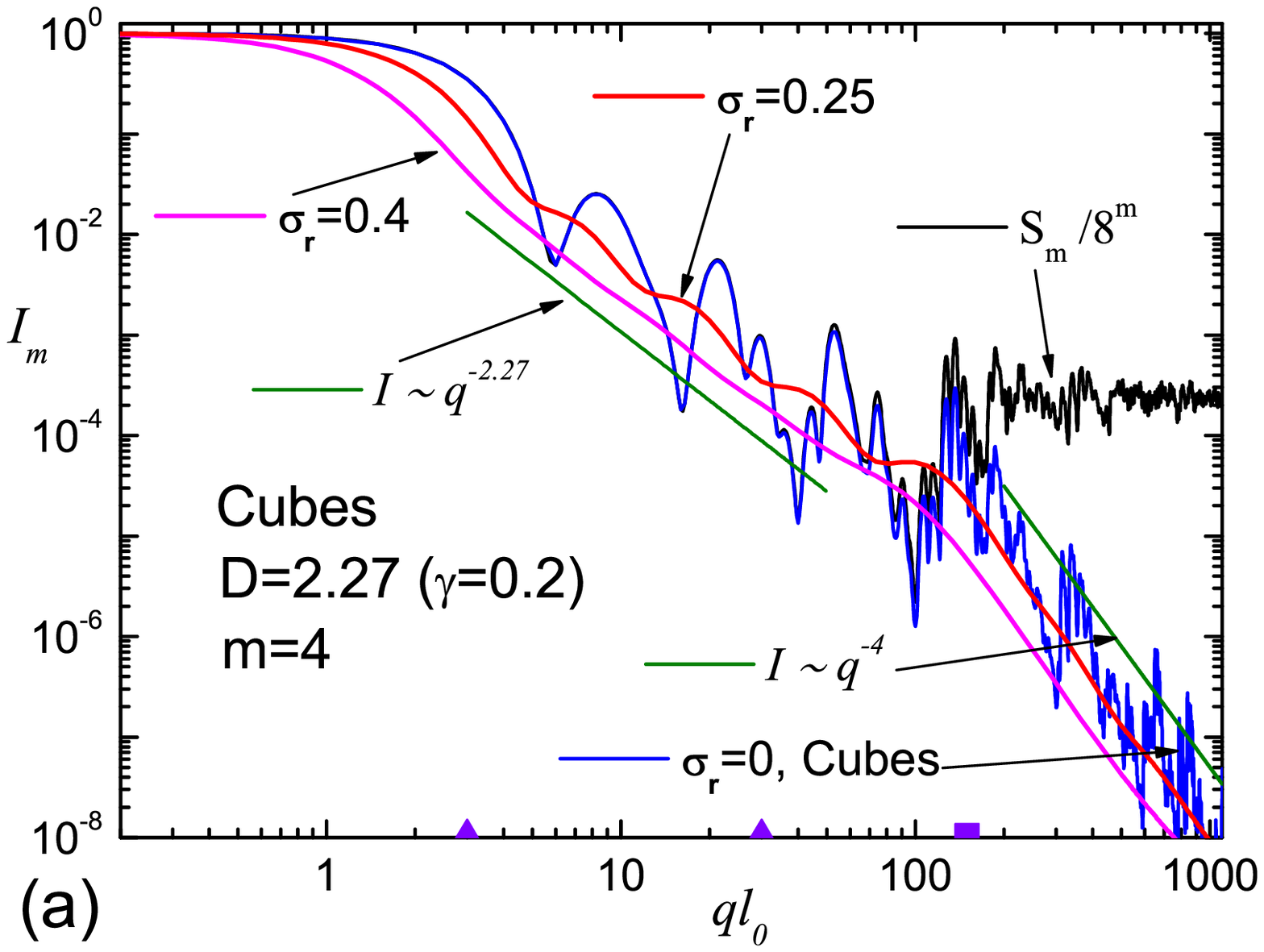}\\
\includegraphics[width=.95\columnwidth]{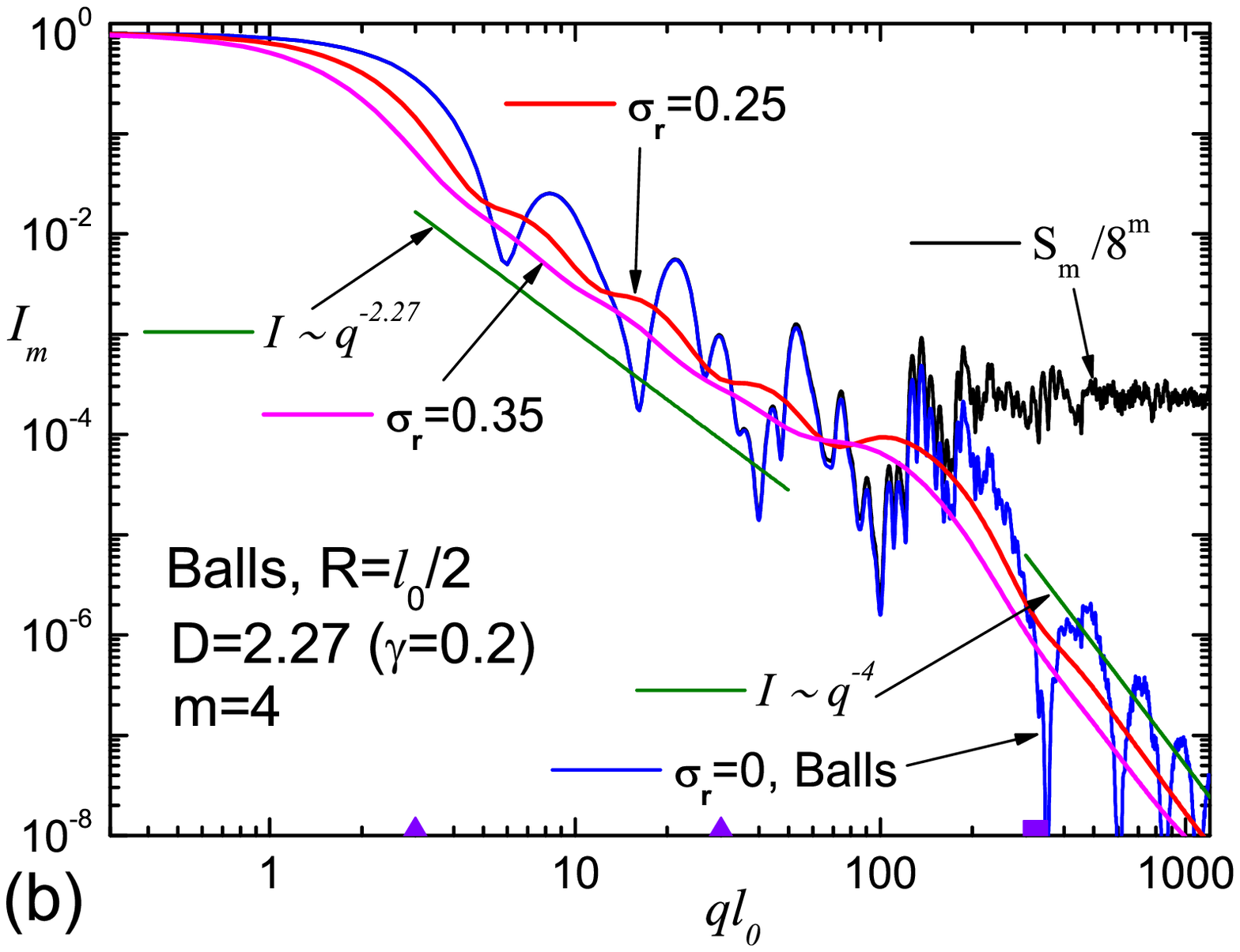}\\
\includegraphics[width=.95\columnwidth]{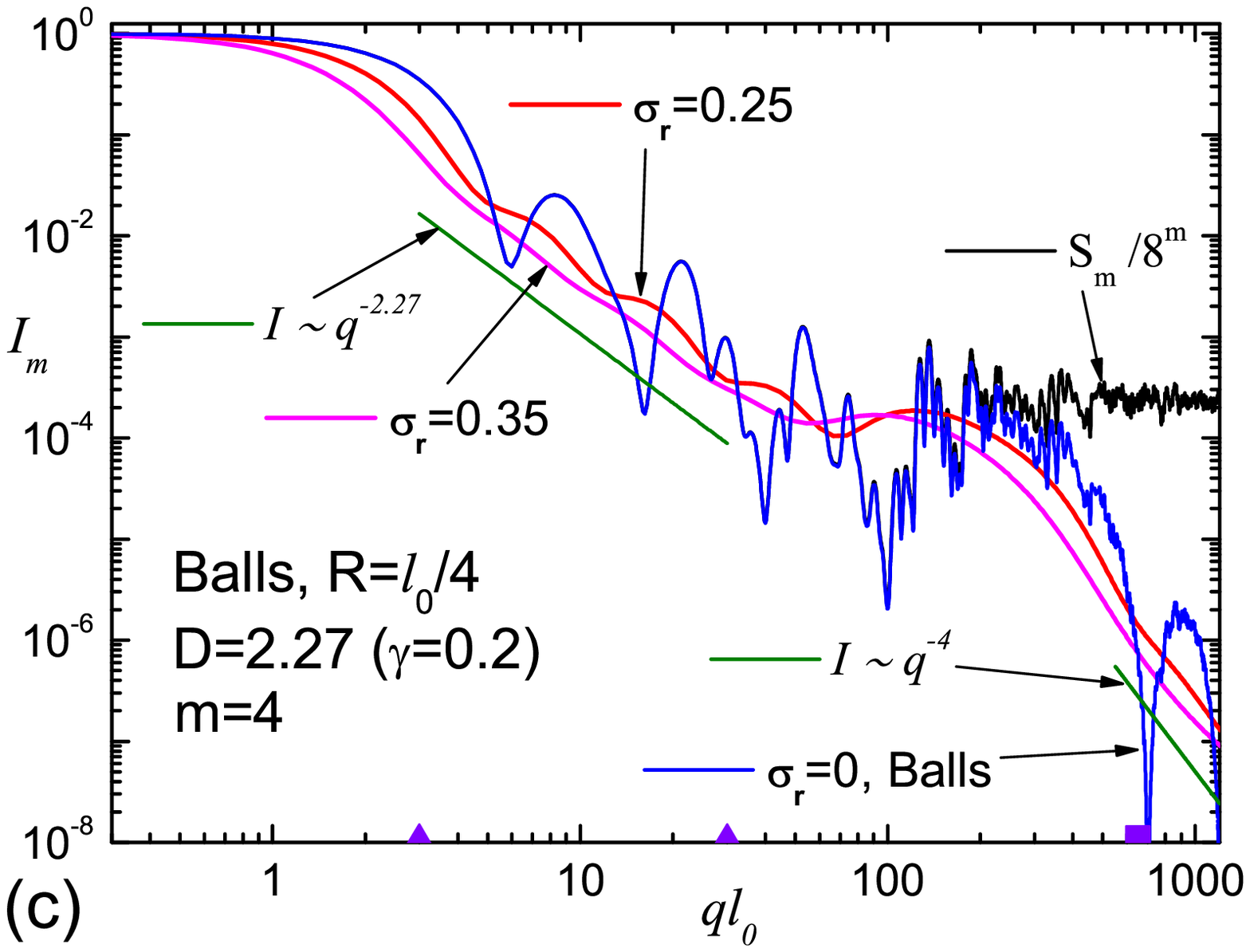}
\caption{Influence of polydispersity on scattering from Cantor fractals of (a) cubes 
and (b)  and (c) balls. The radius of the initial ball (initiator) is $R=l_0/2$ in part (b) and 
$R=l_0/4$ in part (c). The upper monodispersive curve is the structure factor [equation~(\ref{struct})], which in the fractal region [equation~(\ref{powerlaw})] is proportional to the exact intensity [equation~(\ref{ifracreg})]. The borders of the fractal region are shown with filled (violet) triangles on the horizontal axes. In this region, the exponent of the polydisperse curve equals the fractal dimension, while at large momentum $qR\gg 1/\beta_\mathrm{s}^m$, the power law exponent changes from the fractal exponent to four, given by the usual Porod law. The beginning of the Porod region is shown by the filled (violet) squares.}
\label{fig:poly}
\end{figure}

Beyond the fractal region, when the inequality of equation~(\ref{beyondfr}) is satisfied, we derive from equations~(\ref{intapp}) and (\ref{sqasymp})
\begin{equation}
I_m(q l_0)\simeq \langle |F_{0}(\beta_\mathrm{s}^m\bm{q})|^2 \rangle/8^m.
\label{ibeyondfr}
\end{equation}
It follows that, beyond the fractal region, the scattering intensity resembles the 
intensity of the initiator, \emph{i.e.} a cube or ball in the present case (Fig.~\ref{fig:poly}). 
In particular, the maxima of the curve obey Porod's law \cite{glatter82:book}.

\subsubsection{The radius of gyration}

One can calculate the radius of gyration $R_\mathrm{g}$ of the fractal. By definition 
\cite{svergun87:book}, in the Guinier region $I(q)=I(0)(1-q^2R_\mathrm{g}^2/3+\ldots)$. 
Expanding equation~(\ref{fm}) in a power series in $q$ and substituting the result into 
equation~(\ref{final}) yield
\begin{equation}
R_\mathrm{g}=\sqrt{\beta_\mathrm{s}^{2m}R_{\mathrm{g}0}^2
+3\beta_\mathrm{t}^2\frac{1-\beta_\mathrm{s}^{2m}}{1-\beta_\mathrm{s}^2}l_0^2},
\label{rg}
\end{equation}
where $R_{\mathrm{g}0}=l_0/2$ for a uniform cube and $R_{\mathrm{g}0}=R\sqrt{3/5}$ for a 
uniform ball. The limit $m\to\infty$ gives the radius of gyration of the ideal Cantor fractal
\begin{equation}
R_\mathrm{g}=\frac{\sqrt{3}\beta_\mathrm{t}l_0}{\sqrt{1-\beta_\mathrm{s}^2}}. 
\label{rglim}
\end{equation}
As expected, the characteristics of the ideal fractal do not depend on specific shapes from which the fractal is constructed.

\section{Form factor of polydisperse sets}
\label{sec:poly}

Here we study a specific kind of polydispersity, when the whole fractal is 
proportionally scaled and the scales are distributed. This implies that it is sufficient 
to consider the distribution of the fractal length in the above formulae. Then the 
distribution function $D_{N}(l)$ of the scatterer sizes can be considered in such a way 
that $D_{N}(l)dl$ gives the probability of finding a fractal whose size falls within the 
interval $(l,l+dl)$. The intensity can then be written as
\begin{equation}
I_m^{\rm poly}(q)=\frac{\int_{0}^{\infty} \langle F_{m}^{2}(\bm{q}) \rangle 
V_m^{2}(l) D_{N}(l)\,dl} {\int_{0}^{\infty} V_m^{2}(l) D_{N}(l)\,dl}, \label{polyint}
\end{equation}
where $V_m$ is the total volume of the $m$th approximant,
\begin{equation}
V_m=l^3 (2\beta_{\rm s})^{3m}.
\label{volm}
\end{equation}

The influence of particle size distribution on the intensity scattering curves is 
essentially determined by its breadth.  Narrow distributions (\emph{i.e.} log-normal, Schulz, 
Gauss) have no influence on the scattering exponent but control a smoothing of the 
intensity curve, reproducing the power law behaviour for a certain value of the 
variance, while broad ones can change even the scattering exponent \cite{martin86}.

In order to study this influence on scattering from the Cantor sets, we consider a 
log-normal distribution given by
\begin{align}
D_{N}(l)&=\frac{1}{\sigma l \sqrt{2 \pi}}
\exp\left[-\frac{\big(\ln (l/l_0)+\sigma^2/2\big)^{2}}{2 \sigma^2}\right],\label{lognormal}\\
\sigma&\equiv\sqrt{\ln(1+\sigma^2_\mathrm{r})}.\nonumber
\end{align}
Here, $l_0$ and $\sigma_\mathrm{r}$ are the mean length 
and its relative variance, respectively, \emph{i.e.}
\begin{equation}
l_0\equiv\langle l\rangle_D,\quad\sigma_\mathrm{r}\equiv \sqrt{\langle 
l^2\rangle_D-l_0^2}/l_0,
\label{sigmar}
\end{equation}
where $\langle\ldots\rangle_D\equiv\int_0^\infty\ldots D_{N}(l)\,dl$. One can see that 
the relative variance regulates the width of the distribution.

Fig. \ref{fig:poly} represents the scattering curves from mono- and 
polydisperse Cantor sets for the fourth iteration at $\gamma = 0.2$. The curves 
corresponding to the polydisperse cases approach the monodisperse ones as the distribution 
width $\sigma_\mathrm{r}$ tends to zero. Smoothing of intensity curve [equation~(\ref{polyint})] 
increases when the width of the distribution $\sigma_\mathrm{r}$ becomes larger. The most 
interesting effect is changing the power law exponent from the fractal dimension $D$ in the
fractal region [equation~(\ref{powerlaw})] to the usual Porod exponent of four beyond this region. 
Here the intensity is approximately proportional to the scaled intensity of the initiator, 
according to equation~(\ref{ibeyondfr}), thus giving Porod's law.

One can see the well in the intermediate region $ql_0\sim 1/\beta_\mathrm{s}^{m}$, where 
the monodisperse curve is shifted up to the asymptote $1/8^m$. The well becomes even 
more pronounced when the size of the initiator becomes smaller, compare Fig.~\ref{fig:poly}(b) 
and (c).  For a very small initiator size, a plateau should appear here. 
The reason is that the border of the Porod region shifts forward, and until it reaches this border 
the intensity profile coincides with the smoothed structure factor, which has an 
``infinite" plateau. Such behaviour of the scattering intensity has been observed 
experimentally \cite{lebedev08}. A phenomenological approach has been developed 
by Beaucage (1995)~\cite{beaucage95} for studying various exponents in the scattering intensity.

Note that the appearance of the well or ``shelf" between the fractal and Porod regions 
near the volume $1/N_m$ is a consequence of the general asymptote 1 of the 
fractal structure factor [equation~(\ref{sqasymp})]. Thus, the position of the ``shelf" allows us to estimate the 
volume of particles that form the fractal, provided the total fractal volume is known 
(see discussion in Sec.~\ref{sec:genrem}).  

Expanding equation~(\ref{polyint}) in a power series in $q$ and integrating over the 
distribution, we arrive at the radius of gyration given by equations~(\ref{rg}) or 
(\ref{rglim}) multiplied by the factor $(1+\sigma^2_\mathrm{r})^{13/2}$.

\section{Conclusions}
\label{sec:concl}

We consider a model that generalizes the well known one-dimensional Cantor set. It is 
characterized by a scaling factor controlling the fractal dimension, which 
can be varied from zero to three. This is beyond the standard for mass fractals from an 
experimental point of view~\cite{malcai97}, since fractal dimensions less than 1 have not 
yet been observed. The form factor of the generalized Cantor set is calculated 
analytically for arbitrary iteration. This allows us to evaluate the scattering 
intensity for mono- and polydisperse fractals by means of simple integrals. We 
find the values of the asymptotes of the fractal structure factor and reveal their nature. 
The radius of gyration is obtained analytically as well.

The suggested model describes changing the power law exponent from the fractal 
dimension to the usual Porod exponent beyond the fractal region. In the intermediate 
region a typical well or ``shelf" arises, which can be observed experimentally 
\cite{lebedev08}. In comparison with Schmidt and Dacai (1986)~\cite{schmidt86}, we not only calculate the 
scattering intensity for large momenta but also explain its behaviour. Moreover, if an 
experiment observes the threshold between the fractal and Porod regions, the approach suggested here yields the number of particles in the fractal set. Another way to describe the 
experimental data for fractals is \emph{ab initio} methods, see e.g.~\cite{ozerin06}. 

In a number of cases we have \emph{a priori} information about the fractal structure, 
for instance when the fractal synthesis is well controlled by chemical methods. One can then build a fractal model with many parameters and use this model to obtain  additional information from the scattering data. The generalized Cantor set is an 
example of such a construction. A similar scheme can be developed by the same method 
for other types of fractal sets. 

The model can be extended to match specific needs in various ways, including different 
values of the scaling factor at each iteration, and different probability distributions 
for the fractal length and the shapes from which it is constructed.

\acknowledgements

The authors are grateful to A.~N. Ozerin and V.~I. Gordeliy for fruitful discussions.
The work was supported by Russian state contract No. 02.740.11.0542, a grant of Romanian Plenipotentiary Representative at JINR, and the projects JINR--IFIN-HH.


\end{document}